# On the Nature of Lead(II) "Lone Pair"

Christophe Goulaouen[1]* and Jean-Philip Piquemal*[2,3,4]

1- Laboratoire de Chimie Quantique, UMR7177 CNRS et Université de Strasbourg, Strasbourg, France.
2- Laboratoire de Chimie Théorique, Sorbonne Université, UMR7616 CNRS, Paris, France.
3- Institut Universitaire de France, Paris, France.
4- Department of Biomedical Engineering, the University of Texas at Austin, Austin, TX, USA.

**Corresponding authors:**
gourlaouen@unistra.fr; jean-philip.piquemal@sorbonne-université.fr

**Abstract**

We study the quantum chemical nature of the Lead(II) valence basins, sometime called the Lead "lone pair". Using various chemical interpretation tools such as the molecular orbital analysis, Natural Bond Orbitals (NBO), Natural Population Analysis (NPA) and Electron Localization Function (ELF) topological analysis, we study a variety of Lead(II) complexes. A careful analysis of the results show that the optimal structures of the lead complexes are only govern by the 6s and 6p subshells whereas no involvement of the 5d orbitals is found. Similarly, we do not find any significant contribution of the 6d. Therefore, the Pb(II) complexation with its ligand can be explained through the interaction of the $6s^2$ electrons and the accepting 6p orbitals.

We detail the potential structural and dynamical consequences of such electronic structure organization of the Pb (II) valence domain.

## I Introduction

Since the Antiquity, Lead has been considered as a metal of prime interest. However, we now know more about its toxicity but many industrial applications still use it and environmental impacts are large.[1] For this reason, beyond the constant interest for a deeper understanding of the synthetic chemistry and associated properties of Pb(II) compounds, a strong motivation exists towards designing molecules capable to selectively chelate Pb(II) towards applications in Medicine[2,3] or in Environment.[4] If many efforts were dedicated towards the analysis of solid-state properties of lead-containing materials[5,6], the fine comprehension of the Pb(II) coordination at a molecular level remains less documented[7,8,9,10] and theoretical computations have been shown to be important. Indeed, since the first calculation by Shimoni-Livni[11] which highlighted the possibility of holodirected and hemidirected geometries for complexes where Pb(II) is coordinated to four ligands, the origin of the lacuna in low coordinated lead complex has been discussed. Its origin has been linked to the orbitals involved in the complexation. The electronic configuration of the Lead(II) cation (i.e. $Pb^{2+}$) is [Xe] $4f^{14}$ $5d^{10}$ $6s^2$ $6p^0$ $6d^0$. The highest occupied orbital is the full 6s and the lowest unoccupied the empty 6p. The s nature of the HOMO orbital theoretically does not allow any directionality of the electronic pair, so the origin of the lacuna is still under discussion. The first hypothesis was that the s electronic pair was polarized by the 6p orbital that would enable to create a hybrid "sp" lone pair. However, the nature of the orbital and the Natural bond orbital (NBO) analysis show that the p character of the electronic pair is very weak, being about only 5%.[11,12] Understanding this bonding pattern is very important as the tendency of the Lead(II) cation to exhibit a lacuna in its coordination sphere seems strongly linked to the cation's toxicity (i.e. the well-known Lead poisoning). We have shown in a previous work[13] that what we called the "Lead lone pair" was at the origin of perturbation of proteins through its ability to restructure the metallic coordination sphere.

In a previous paper dedicated to series of Lead ligands, we have studied by mean of the Electron Localization Function (ELF) topological analysis the evolution of the volume and the density of Lead(II) valence basin.[14] We obtained a population larger than two showing that charge transfer could play a role in the emergence of a directional valence basin. However, all calculations were performed using large core pseudopotentials in which the 5d electrons were merged in the core so that only the 6s electrons remain. Therefore, we were not able to investigate the possible role of these low lying 5d orbitals. In this contribution we propose to come back to the subject using all the available analytic tools to investigate the nature of the Lead(II) cation valence basin cation in order to determine the orbital involved within it. Then, we will extend our study to the flexibility of the hemidirected structures of lead complexes.

### II Computational details

A first set of calculations have been performed using the TURBOMOLE package[15] at MP2 the level of theory using a def2-TZVP basis set[16] (with associated small core pseudopotential for Pb and I atoms) in gas phase. The quantum chemical results have been analyzed by means of Natural Population Analyses (NPA) and Natural Bonding Orbital (NBO) analyses.[17] To perform the topological analyses, we extracted the TURBOMOLE optimized geometries and computed the wavefunction with GAUSSIAN09 version D01 package[18] within the B3LYP[19] formalism with the same basis set.

On the basis of this wavefunction, ELF (Electron Localization Function) calculations and the topological analysis of the ELF functions, together with specific integrations, have been performed using the TopMod package.[20,21,22,23,24,25,26,27] We here just recall that within the framework of the topological analysis of the ELF function, space is partitioned into *basins*, each of them having a chemical meaning. Such basins are classified as: i) core basins surrounding nuclei, ii) valence basins characterized by their synaptic order.[21] Further details can be found in the above-mentioned references.

It has been shown possible to extend the ELF approach to pseudopotential approaches.[28] Using small-core pseudopotentials provides the so-called *semi-external cores* and allows to determine the synapticity of well-defined valence basins.[29] Using large-core pseudopotentials preserves the number and the properties of valence basins. Only small core pseudo potential have been used. In the present contribution, we will focus on V(Pb), the valence monosynaptic basin associated to the valence electrons of Pb(II). For a given complex [Pb(II)L$_n$]$^q$, we will use the following notations: V(Pb) is the ELF basin defined previously, N(Pb) and ω(Pb) are respectively the population and the volume associated to this basin. A last parameter is the distance d(Pb) between the ELF attractor of lead valence basin V(Pb) and the lead cation itself

A second set of geometry optimizations have been performed with with GAUSSIAN at DFT level of theory with B3LYP and the more recent ωB97XD[30] functionals on the [Pb(CN)$_3$(HCl)]$^-$, [Pb(CN)$_3$(ClH)]$^-$, [Pb(CO)$_3$(HCl)]$^{2+}$ and [Pb(CO)$_3$(ClH)]$^{2+}$ complexes. To explore the nature of the interaction between the Pb(II) complexes and the extra HCl probe molecule, a NCI (non-covalent interactions) study has been performed on the basis of the optimized geometry wavefunction using the NCIPLOT package[31] (nciplot3 version). For some complexes, we also searched for the transition state to compute the interconversion barriers. All the provided energies are Gibbs Free energies as obtained from GAUSSIAN frequency calculation at 298 K.

We want to point out, that for ligands in the first coordination sphere, there are very few observed differences for the bond lengths between the different methods. In practice, the values are closed to those already published as illustrated by the Table S1 ESI.

### III Results and discussion

We performed full geometry optimization on several Lead complexes. First is the $[Pb(CO)_n]^{2+}$ series, which is easy to analyze thanks to ligand rigidity. Second come complexes with various organic or inorganic ligands. All these complexes are a small sample of the potentially existing Lead complexes however they have all been studied and represent a variety of coordination modes and conformations. On all these complexes, we have studied the involvement of Lead orbitals in the coordination and on the complex topology.

Pb(II) electronic configuration is [Xe] $4f^{14}$ $5d^{10}$ $6s^2$ $6p^0$. The 4f shells are very deep, their absolute energies being about -3600 kcal/mol ($4f_{7/2}$) and -3720 kcal/mol ($4f_{5/2}$)[32]. From a chemical point of view, they are considered as inert and are included in the core potential. The 5d orbitals present a much higher energetic: -650 kcal/mol ($5d_{5/2}$) and -714 kcal/mol ($5d_{3/2}$). However, they remain low in energy, back-donation or electron promotion from these orbitals was checked: they are directional and can mix efficiently with the ligand p orbitals as frequently observed in transition metal complexes.[33] The population of full 5d orbitals or empty 6d orbital was checked through NBO analysis (from the MP2 calculations) for the whole series of complexes studied. The 5d orbitals are weakly depopulated and this depopulation seems to be independent from the ligand: there are 9.87 electrons in these orbitals in $[Pb(H)_3]^-$ complex. The value is the same (9.87 electrons) in $[Pb(CN)_3]^-$ complex and in $[Pb(OH_2)_3]^{2+}$ complex (9.89 electrons). Charge transfer towards the empty 6d is also very weak from 0.05 electrons in $[Pb(H)_3]^-$ to only 0.02 electrons in $[Pb(OH_2)_3]^{2+}$ complex. The occupation of these orbitals almost does not differ from that of the isolated $Pb^{2+}$ cation and is furthermore almost ligand independent. As a conclusion, these orbitals are not involved in the bonding and will not be discussed any further. The true valence orbitals are therefore the full $6s_{1/2}$ (-355 kcal/mol) and the empty 6p orbitals: -138 kcal/mol ($6p_{3/2}$) and -172 kcal/mol ($6p_{1/2}$) that will be the focus of our study.

These two sets of orbitals are the most important as the 6p will accept electrons from donating ligands and the full 6s orbital thanks to its diffuseness will prevent a too short approach of the ligand. Their role has been discussed[11] and it has been shown that in the presence of a ligand field the external valence shell of the cation distorted to allow a close contact with the ligands. Therefore at this point the questions of the precise nature of the Lead(II) valence basin and of the concrete structural and dynamical consequences of its existence remain to be addressed.

## III.1 Structures and properties of lead complexes

Different sets of Pb(II) complexes were investigated to test the influence of several parameters on the geometries. First, we recall the results for the $[Pb(CO)_n]^{2+}$ series (n=1 to 6) which topology will be explored further. Then we compare the structures of $[Pb(X)_3]^-$ complexes (X = HO$^-$, CN$^-$, HS$^-$, H$^-$, Cl$^-$) to understand the influence of the anion on the structure. On the $[Pb(CO)_n]^{2+}$ series, we retrieve the results previously published[34]: from n=1 to n=3, the Pb-C distances are insensitive to n (see Table 1). We observe a small depopulation of the 6s orbital (less than 0.1 electrons) and a steady increase of the 6p electron population. However, it should be noticed that this 6s depopulation increases from n=1 to n=3 and is almost nil for higher coordination numbers. This could be linked to the presence of close carbonyl that destabilize the 6s orbital to ensure coordination. Starting from coordination number 4, the Pb-C distances are no longer identical. In $[Pb(CO)_4]^{2+}$ complex, there are two short distances and two longer. The latter corresponding to the two CO being in trans position and the former to CO for which there are no ligand in trans position. As the lengthening of the Pb-C distances diminishes the interactions, the 6s population is allowed to increase whereas the 6p population exhibit a more modest increase as the long Pb-C distances disfavor the ligand donation. The general shape of the complexes is a butterfly for n=4, square based pyramid for n=5, the lead cation being outside the square plane and a perfect octahedron for n=6. Interestingly, even for n=2, the structure deviates from the ideal 90° for the C-Pb-C angle, the value is smaller, despite the ligand-ligand repulsion showing that it is not this interaction that governs the structure.

|  | Pb-C | C-Pb-C | 6s | 6p | V(Pb) | | |
|---|---|---|---|---|---|---|---|
|  |  |  |  |  | ω(Pb) | N(Pb) | d(Pb) |
| $[Pb(CO)]^{2+}$ | 2.612 | NC | 1.94 | 0.17 | 165.6 | 1.25 | 1.583 |
| $[Pb(CO)_2]^{2+}$ | 2.627 | 83.1 | 1.93 | 0.35 | 160.6 | 1.42 | 1.587 |
| $[Pb(CO)_3]^{2+}$ | 2.635 | 82.1 | 1.91 | 0.58 | 156.0 | 1.52 | 1.620 |
| $[Pb(CO)_4]^{2+}$ | 2.635 2.777 | 81.5 150.0 | 1.95 | 0.59 | 135.3 | 1.44 | 1.616 |
| $[Pb(CO)_5]^{2+}$ | 2.635 2.796 | 76.3 86.6 152.5 | 1.95 | 0.71 | 106.5 | 1.27 | 1.618 |
| $[Pb(CO)_6]^{2+}$ | 2.845 | 90.0 | 1.98 | 0.77 |  |  |  |

**Table 1:** Properties of the $[Pb(CO)_n]^{2+}$ complexes. Pb-C bond lengths in Angstroms. C-Pb-C angle in degrees. Electronic population of the 6s and 6p shells of lead from NBO analysis. Volume (ω(Pb) in Å$^3$),

population (N(Pb) in electrons) of lead valence basin and distance (d(Pb) in Å) of the V(Pb) ELF attractor and cation.

The second series of complexes studied is $[Pb(Cl)_n]^q$ with n = 1 to 4 (Figure 1). All attempts to optimize higher coordinated structure led to ligand decoordination. The Pb-Cl distances are shorter than the Pb-C distances (see Table 2) and we can link this to the slightly larger 6s depopulation and 6p population. The geometry of the complexes has a strong effect on the bond length and orbital population. For $[Pb(Cl)_2]$ and $[Pb(Cl)_3]^-$ complexes, the structures of highest symmetries ($D_{\infty h}$ and $D_{3h}$ respectively) lead to strong Pb-Cl distance lengthening (0.2 and 0.1 Å respectively) compared to their lower symmetry analogues. Simultaneously, the repopulation of the 6s shell and drop of the 6p shells is observed, although it is less pronounced in the $[Pb(Cl)_3]^-$ complex, highlighting again the strong trans effect in lead complexes. This is well illustrated by the characteristics of the $[Pb(Cl)_4]^{2-}$ complex. Contrary to the $[Pb(CO)_4]^{2+}$ complex, the structure of $[Pb(Cl)_4]^{2-}$ has the highest $T_d$ symmetry. The addition of a fourth chloride anion on the $[Pb(Cl)_3]^-$ complex leads to a strong increase of the Pb-Cl distance. In addition, a drop of the 6p shell population is observed despite the presence of an extra anion and also a repopulation of the 6s shell. The symmetric structure of the complex and the bond length increase by diminishing the overlap with the 6p orbitals and lowering the constraint exert of the 6s shell explain these electronic evolutions.

|  | Pb-Cl | Cl-Pb-Cl | 6s | 6p | V(Pb) | | |
|---|---|---|---|---|---|---|---|
|  |  |  |  |  | ω(Pb) | N(Pb) | d(Pb) |
| $[Pb(Cl)]^+$ | 2.348 | NC | 1.91 | 0.55 | 216.4 | 1.78 | 1.707 |
| $[Pb(Cl)_2]$ ($C_{2v}$) | 2.449 | 99.3 | 1.87 | 0.89 | 213.2 | 1.95 | 1.787 |
| $[Pb(Cl)_2]$ ($D_{\infty h}$) | 2.566 | 180 | 1.99 | 0.39 | 182.7 | 1.32 | 1.401 |
| $[Pb(Cl)_3]^-$ ($C_{3v}$) | 2.560 | 99.8 | 1.84 | 0.96 | 186.6 | 1.94 | 1.837 |
| $[Pb(Cl)_3]^-$ ($D_{3h}$) | 2.660 | 120.0 | 1.93 | 0.72 | 79.48 | 0.64 | 1.439 |
| $[Pb(Cl)_4]^{2-}$ ($T_d$) | 2.781 | 109.5 | 1.93 | 0.74 | 56.9 | 0.62 | 1.225 |

Table 2 : Properties of the $[Pb(Cl)_n]^q$ complexes. Pb-Cl bond lengths in Angstroms. Cl-Pb-Cl angle in degrees. Electronic population of the 6s and 6p shells of lead from NBO analysis. Volume (ω(Pb) in Å³), population (N(Pb) in electrons) of lead valence basin and distance (d(Pb) in Å) of the V(Pb) ELF attractor and cation.

| [Pb(Cl)]$^+$ $D_{\infty h}$ | [Pb(Cl)$_2$] $C_{2v}$ | [Pb(Cl)$_2$] $D_{\infty h}$ |
|---|---|---|
| 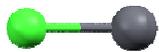 | 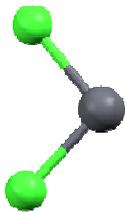 | 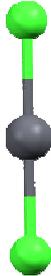 |
| 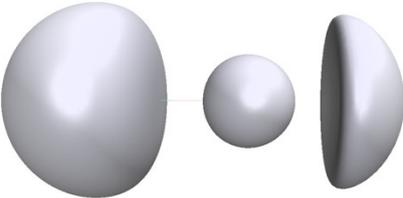 | 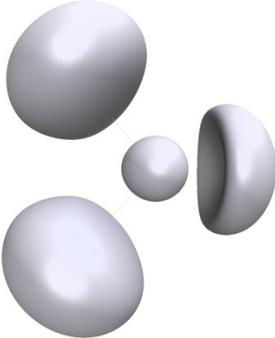 | 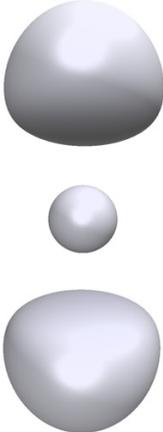 |
| 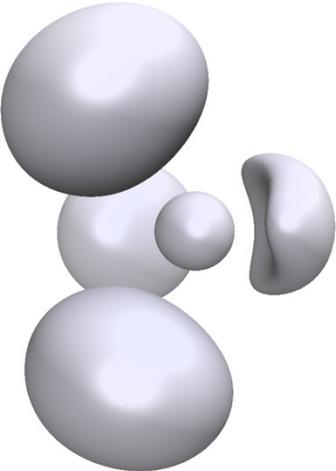 | 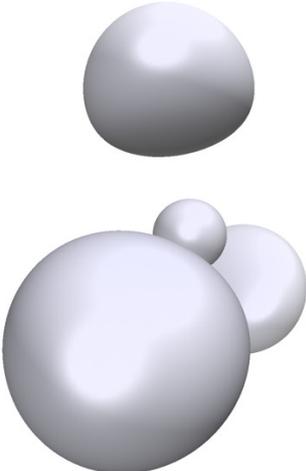 | 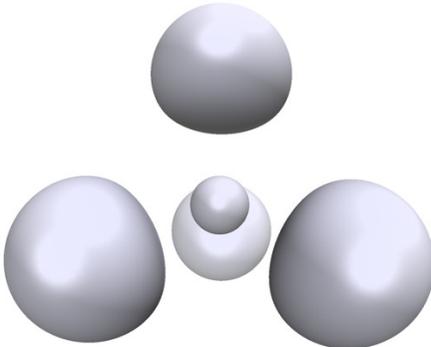 |
| 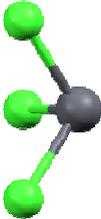 | 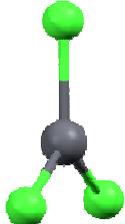 | 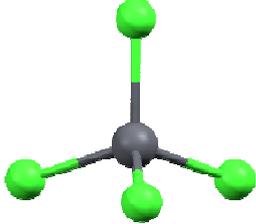 |
| [Pb(Cl)$_3$]$^-$ $C_{3v}$ | [Pb(Cl)$_3$]$^-$ $D_{3h}$ | [Pb(Cl)$_4$]$^{2-}$ $T_d$ |

**Figure 1 : Structures and ELF function ($\eta$=0.65) of the different [Pb(Cl)$_n$]$^q$ complexes.**

Finally, we compare the structure of a series of Pb(L)$_3$ complexes, either dicationic or monoanionic (Table 3). Surprisingly, no clear trends appear on the bond Pb-L bond lengths, L-Pb-L angles or orbital occupations. Indeed, if the Pb-L distances are generally shorter if L is anionic (L = HO$^-$) than if L is neutral (L = H$_2$O). However, the Pb-L distances are almost identical for L = H$_2$O and L = H$_3$C$^-$. Things are clearer for the L-Pb-L angle. It is smaller for the neutral ligands, between 80 to 90°, than for the anions for which it values around 90-92°. This last value can be understood as it maximizes the overlap with the cation 6p orbitals. However, the halide complexes deviate from this value, the angle being larger than 97° and increasing with the halide size (from F$^-$ to I$^-$). The 6s orbital depopulation is stronger for the anionic ligand (up to 0.34 electron for [Pb(H)$_3$]$^-$) than for the neutral one for which it is smaller than 0.1 electron. The population of the 6p orbital is higher for the anionic ligand, due to their better σ-donation ability and the 6s to 6p electron promotion than the neutral ligand. Inside the anionic ligand series, soft ligand according to HSAB theory favors greater 6p population (see halide series or compare L = HO$^-$ and L = HS$^-$).

|  | Pb-L | L-Pb-L | 6s | 6p | V(Pb) | | |
|---|---|---|---|---|---|---|---|
|  |  |  |  |  | ω(Pb) | N(Pb) | d(Pb) |
| [Pb(H)$_3$]$^-$ | 1.848 | 91.2 | 1.66 | 2.45 | 286.5 | 2.51 | 1.966 |
| [Pb(Me)$_3$]$^-$ | 2.328 | 90.8 | 1.68 | 1.80 | 282.9 | 2.48 | 1.988 |
| [Pb(F)$_3$]$^-$ | 2.115 | 97.1 | 1.78 | 0.61 | 210.9 | 2.01 | 1.898 |
| [Pb(Cl)$_3$]$^-$ | 2.560 | 99.8 | 1.84 | 0.96 | 186.6 | 1.94 | 1.837 |
| [Pb(Br)$_3$]$^-$ | 2.719 | 100.6 | 1.87 | 1.15 | 183.5 | 1.96 | 1.810 |
| [Pb(I)$_3$]$^-$ | 2.911 | 100.8 | 1.88 | 1.34 | 182.6 | 2.04 | 1.787 |
| [Pb(CN)$_3$]$^-$ | 2.306 | 91.9 | 1.73 | 1.50 | 212.5 | 2.19 | 1.882 |
| [Pb(OH)$_3$]$^-$ | 2.176 | 91.8 | 1.78 | 0.96 | 236.7 | 2.13 | 1.935 |
| [Pb(SH)$_3$]$^-$ | 2.636 | 89.9 | 1.80 | 1.34 | 214.4 | 2.17 | 1.859 |
| [Pb(HCN)$_3$]$^{2+}$ | 2.415 | 84.7 | 1.89 | 0.43 | 161.8 | 1.65 | 1.682 |
| [Pb(CO)$_3$]$^{2+}$ | 2.635 | 82.1 | 1.91 | 0.58 | 156.0 | 1.52 | 1.620 |
| [Pb(OH$_2$)$_3$]$^{2+}$ | 2.374 | 83.8 | 1.91 | 0.31 | 158.9 | 1.58 | 1.638 |
| [Pb(NH$_3$)$_3$]$^{2+}$ | 2.451 | 90.0 | 1.89 | 0.61 | 172.9 | 1.82 | 1.690 |

Table 3: Properties of [Pb(L)$_3$]$^q$ complexes. Pb-L bond length in Angströms. L-Pb-L angle in degrees. Electronic population of the 6s and 6p shells of lead from NBO analysis. Volume (ω(Pb) in Å$^3$), population (N(Pb) in electrons) of lead valence basin and distance (d(Pb) in Å) of the V(Pb) ELF attractor and cation.

One last point concerns the heterogeneous complexes. From the $[Pb(CO)_n]^{2+}$ and $[Pb(Cl)_n]^q$ series, we showed that putting ligands in trans positions lead to strong Pb-L distance lengthening. Stable structures can be optimized up to n = 6 for the CO series but up to only n = 4 for the Cl series. In experimental conditions (biological environment and/or solvated media), the lead coordination sphere is generally heterogeneous with both neutral and anionic ligand. To explore the stability of such edifices, we have progressively replaced one CO ligand within the optimized $[Pb(CO)_6]^{2+}$ structure by $CN^-$ and reoptimized the structure. The structure of the $[Pb(CN)(CO)_5]^+$ complex is highly distorted compared to the $[Pb(CO)_6]^{2+}$ reference (Figure 2). The CO ligand trans to the cyano one is only weakly bonded to the complex with a Pb-C distance of 3.643 Å. However, the carbonyls located in cis are also affected. If the bond length (2.828 Å) is similar to that in the hexacarbonyl complex, the angle between the cyano and the carbonyl ligand in cis is only of 76.5°.

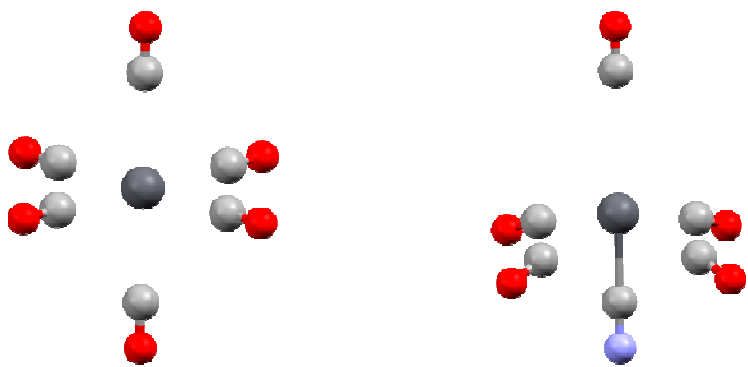

**Figure 2 : Structure of $[Pb(CO)_6]^{2+}$ (left) and $[Pb(CO)_6(CN)]^+$ complexes.**

Two structures are possible for the $[Pb(CN)_2(CO)_4]$ complex. A stable structure was found when the cyano ligands are in trans position. In that case the shape of the complex is that of an octahedron, with two short Pb-CN distance of 2.560 Å though much larger than in the $[Pb(CN)_3]^-$ complex (Table 3), and four long Pb-CO distances (2.911 Å) larger than in the $[Pb(CO)_6]^{2+}$ complex. No stable structure was found when the cyano ligands are in cis position. Indeed, the CO in trans position with the cyano ligands are expelled from the complex. For $[Pb(CN)_3(CO)_3]^-$, the situation is similar, all the carbonyl complexes are expelled from the complex. This suggests that in gas phase, the trans effect is so strong, that no neutral ligand can bind trans to an anionic ligand.

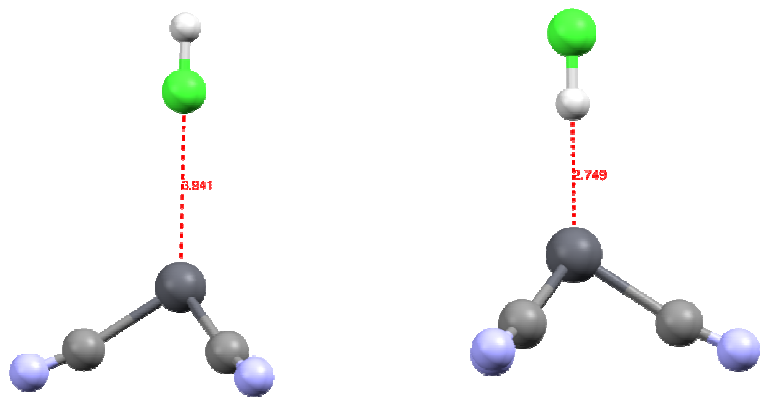

**Figure 3 : Structure of the [Pb(CN)$_3$(ClH)]$^-$ (left) and [Pb(CN)$_3$(HCl)]$^-$ (right) complexes with distances in Å.**

We studied the complexation of an HCl molecule onto the [Pb(CN)$_3$]$^-$ complex. The approach is along the C$_3$ axis of the complex either through the chlorine [Pb(CN)$_3$(ClH)]$^-$ or hydrogen atom [Pb(CN)$_3$(HCl)]$^-$. The [Pb(CN)$_3$(HCl)]$^-$ is the most stable by 4.2 kcal/mol with a Pb-H bond length of 2.749 Å. On the contrary, in the [Pb(CN)$_3$(ClH)]$^-$ complex, the Pb-Cl distance is of 3.941 Å (Figure 3). This suggests that when searching a new ligand to extract lead from a media, not only the direct coordination sphere has to be considered but also a positive pole may play a role to complex this coordination sphere.

### III.3 Topology and nature of interactions

In a previous work, we have shown that the presence of the lead valence controls the structure of lead complexes. Especially, this basin explains the deactivation of a zinc finger metalloenzyme[13] or the structure of various complexes. We were able to correlate its properties (volume and density) to the nature of the ligand and also to put in evidence some unsuspected effect on the ligand like with thiocyanate ligand[14].

The ELF basins structures of Lead are strongly dependent on the cation coordination sphere. As we use a pseudopotential for our calculations, the computed ELF basins only contain the 5s, 5p, 5d, 6s and 6p shells. From the ELF point of view, this generates two kinds of ELF basins, a core basin and a valence basin. The electrons of the fifth shell and part of 6s generate the core one, the valence basin containing the rest of the 6s electrons and the population transferred in the 6p by ligand donation. This can be illustrated by the result performed on an isolated Pb$^{(II)}$ cation: only 0.82 electrons are present in the valence basin, whereas 19.18 electrons are in the core, which forms a perfect sphere around the cation (Figure 4). In the presence of a ligand field, both core and valence basins are affected. The core basin, which for lead represents the most external shell of core electrons in our computational conditions, splits itself into several fragments. This phenomenon is called subvalence and accounts for the cation polarizability[35]. We also observe a drop of the basin electronic population which falls to 18.19 electrons in the $[Pb(CN)_3]^-$ complex. This is another illustration of the 6s to 6p electron promotion.

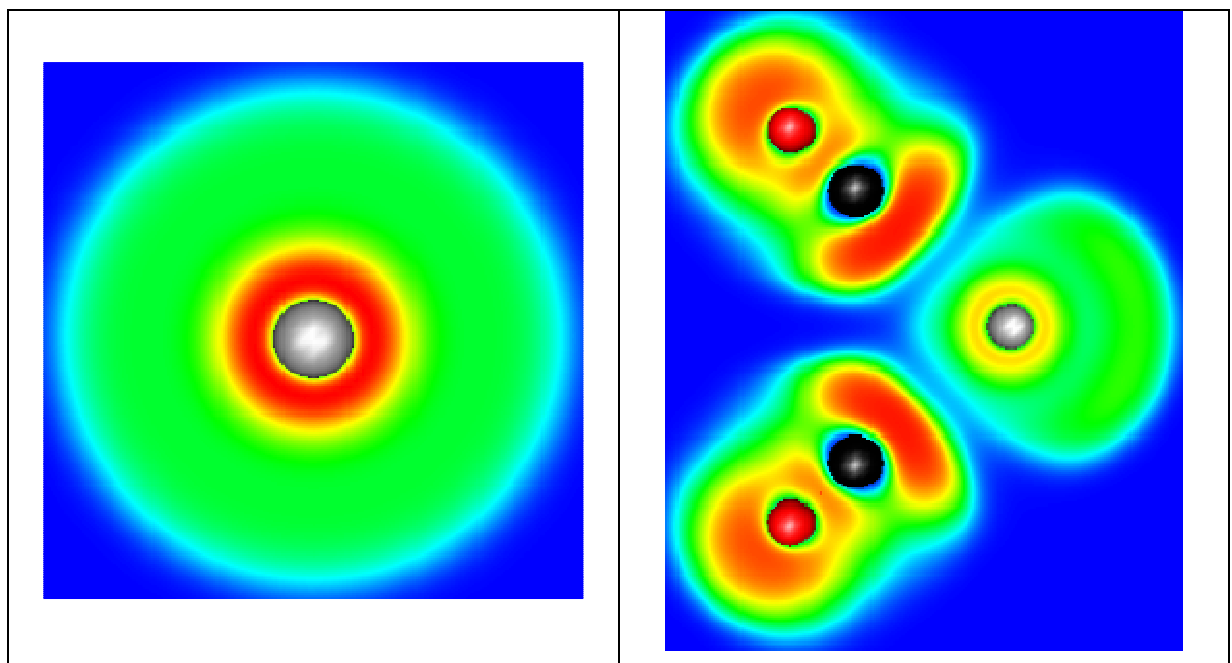

**Figure 4: Cut plane of the ELF function of an isolated $Pb^{2+}$ cation (left) and along the $\sigma_v$ plane of the $[Pb(CO)_2]^{2+}$ complex. The range of the ELF functions varies from 0 (deep blue) to 1 (deep red).**

Upon ligand coordination, the valence basin is highly distorted. When a ligand binds to the cation, the basin population increases due to the 6p population. Furthermore, these orbitals being directional, they allow for the distortion of the basin, which is repelled to the opposite side of the ligand. In a multi-ligated complex, the valence basin tends to localize in the area of the weakest ligand field. This means that for low coordination (n = 1 to 3), the ligand field will adopt a non-spherical (Hemidirected) distribution and the valence basin attractor moves on the opposite side of the ligand. For higher coordination numbers, two kinds ligand distribution are possible depending on the ligand nature. For n=4, the complex $[Pb(Cl)_4]^{2-}$ adopts a spherical distribution (holodirected) whereas the complex $[Pb(CO)_4]^{2+}$ adopts an anisotropic structure (hemidirected). From the complexes studied and the result of Shimoni *et al*[11], it seems that more the ligand is σ-donor, the higher the probability for the structure to be holodirected. It is the case for the halide (except fluoride) or for the hydride and $H_3C^-$. To our knowledge, for n = 5, the complex have only been observed in a hemidirected structure, forming an octahedron in which the valence basin localizes in the lacunary position. For n = 6, the complex is holodirected.

Table 2 illustrates the effect of the ligand field on the valence basin properties. We have compared its volume and population in different structures of the [Pb(Cl)$_2$] and [Pb(Cl)$_3$]$^-$ complexes. Moving from hemidirected to holodirected geometry leads to a significant drop of the volume and electronic population. For [Pb(Cl)$_3$]$^-$, the population decreases from 1.94 electrons ($C_{3v}$) to 0.64 electrons ($D_{3h}$). This evolution is partly due to the smaller available space into which the valence basin can expand. In the hemidirected $C_{3v}$ form, more than half of the coordination sphere is left available for the valence basin whereas in the holodirected $D_{3h}$, the basin has to split into two halves on each side of the complex. This trend and the capability of the basin to expand explain the values of the L-Pb-L angle in Table 3. In the anionic complexes this angle is slightly larger than the ideal 90° value, which maximizes the overlap between the 6p orbitals and the ligand. The small deviation may be due to the inter ligand charge repulsion. On the contrary, for cationic complexes this L-Pb-L angle is significantly lower (except for [Pb(NH$_3$)$_3$]$^{2+}$). As there is less charge repulsion between the ligands, the pressure exerted by the valence basin on the ligand forces the reduction of this angle despite a lower overlap with the 6p orbitals.

Another parameter may be discussed, the position of the ELF attractor(s) and especially its or their distances with the lead cation. Indeed, for most of the complexes, there is only one Pb$^{2+}$ valence basin. The structure of the complexes is a tetrahedron, with V(Pb) occupying one of the summits and the cation being at the center (Figure 5, top right). Though for the smallest donor ligand, namely OH$_2$, NH$_3$ or HCN the valence basin is split in three (Figure 5, top left). For these compounds there are three valence basins. The description of Pb$_{2+}$ valence basins cannot be easily reduced to a hybridized sp$^3$ lone pair, it is much more versatile.

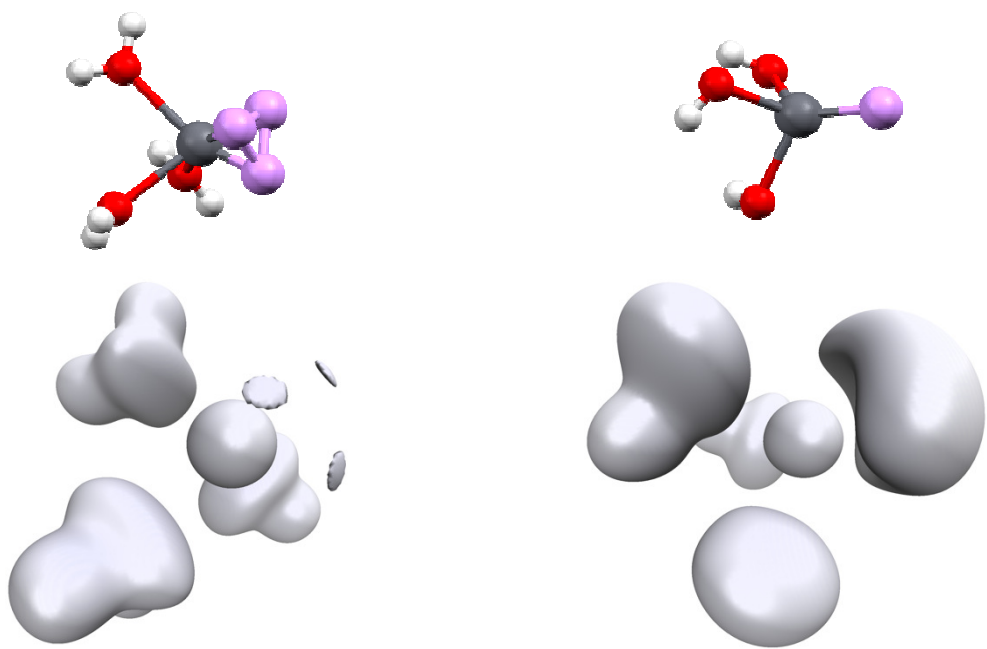

**Figure 5 : Position of the V(Pb) attractors (top) and associated ELF isosurface (η=0.59, bottom) of [Pb(OH$_2$)$_3$]$^{2+}$ (left) and [Pb(OH)$_3$]$^-$ (right) complexes.**

From Table 1 and Table 2, it is shown that this attractor goes further from the cation up to n=3 then its distances does not evolve any longer in the hemidirected structures. This is the effect of the Pb$^{2+}$ polarization due to the ligand field. This is confirmed by Table 3, in which d(Pb), the ELF attractor distances from lead cation, is correlated to the σ-donating forces of the ligand. This force can be assimilated to N(Pb), the population of V(Pb). Plotting d(Pb) *vs.* N(Pb) lead to a linear arrangement (Figure S1, ESI) of the ligand: the more donation to lead there are, the further V(Pb) is repealed.

This valence basin arising from the polarization of the Pb$^{2+}$ external electronic shell traduces an excess of electronic density trans to the more σ-donor ligand. In consequence, there will be a discrimination on the ligand which can bind on the cation: the less donating will be expelled from the coordination sphere when being in trans to a stronger donor group. This is illustrated by the [Pb(CO)$_3$(CN)$_3$]$^-$ complex in which the neutral CO are unstable in the first coordination sphere. This valence basin will also have further consequence on the organization of the cation second coordination sphere. As already mentioned, this valence basin forms an electronic shield preventing further coordination on this position. However, we attempted to approach an HCl molecule on the cation either through the chlorine atom or through the hydrogen atoms. The structure with the hydrogen atom pointing toward the lead (Pb-H distance at 2.781Å) is more stable by 4.3 kcal/mol than the conformer (Pb-Cl distance at 3.941 Å). The NCI analysis performed on the optimized structures (Figure 2) shows the presence of an attractive electrostatic force between the ELF valence basin of the cation and the hydrogen valence basin of the HCl. This traduces that the cation valence basin occupying the complex lacuna can be described as a negative pole which is able to interact with positively charged fragment mimicking a sort of hydrogen bond as suggested by Hancock.[36] Furthermore, it should be noticed that all the structures discussed until now have been optimized at MP2 level of theory. An optimization with B3LYP functional without dispersion correction of the $[Pb(CN)_3(ClH)]^-$ complex failed to find any minima and led to the decoordination of the complex. On the contrary, optimization with ωB97XD functional lead to results close to the MP2 ones. The $[Pb(CN)_3(HCl)]^-$ structure (Pb-H distance of 2.884 Å) is more stable than the $[Pb(CN)_3(ClH)]^-$ one (Pb-Cl distance of 4.324 Å) by 2 kcal/mol. It will be critical in lead complexes to include dispersion corrections (e.g. through Grimme's correction[37]) when the lead valence basin may interact with second coordination sphere ligand. We further explore the importance of the population of $Pb^{2+}$ valence basin by complexing the HCl on the $[Pb(CO)_3]^{2+}$ complex with ωB97XD functional. The situation is completely different. This time, no minima is found for the $[Pb(CO)_3(HCl)]^{2+}$ complex. When the hydrogen points toward the lead, the HCl molecule is expelled from the complex. On the contrary, there is a minima for the $[Pb(CO)_3(ClH)]^{2+}$ complex with a Pb-Cl distance of 3.041 Å, much shorter than in the $[Pb(CN)_3(ClH)]^-$ complex. This suggests that the structuration of the solvent (the orientation of the molecule) around the valence basin of $Pb^{2+}$ will be dependent of the ligand hold by the cation.

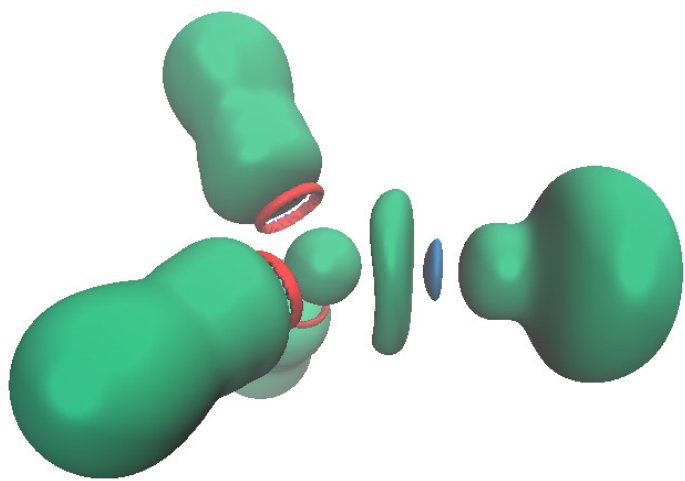

**Figure 6 : Topology of the $[Pb(CN)_3(HCl)]^-$ complex. In green are the ELF basin (isovalue η=0.6), in red and blue the NCI (isovalue η=0.552) interactions. The area in red correspond to steric repulsion and in blue to electrostatic attraction.**

### III.4 Interconversion barriers

The hemidirected character of the $[Pb(L)_3]^q$ complexes (isostructural to $NH_3$) implies the existence of two conformer separated by an interconversion barrier. For some complexes, we have computed the transition state between the two minima and determined this barrier. These calculations have been performed with GAUSSIAN. Surprisingly, the transition state for the interconversion of most complexes is not of $D_{3h}$ but adopt a T-shape (Figure 7). Consequently, the three Pb-L distances are not identical with one short and two long for the two ligands in trans position. Only the mono atomic ligand ($H^-$, $F^-$ and $Cl^-$) exhibits a $D_{3h}$ transition state.

The interconversion barrier is strongly dependent of the ligand (Table 4). The main point seems to be the σ-donor ability. The strongest σ-donor $[Pb(H)_3]^-$ and $[Pb(Me)_3]^-$ have the highest barrier, more than 40 kcal/mol, meaning that the interconversion is not thermally accessible at room temperature. On the contrary, poor σ-donor ligands such as water have very low barrier, and does not have any fixed configuration. Most of the ligand exhibits moderate value, between 15 to 30 kcal/mol. This suggests the possibility to synthesize optically active organolead complexes.

Again, the value of this barrier reflects the flexibility/rigidity of the lead valence basin. Formally the interconversion consists into the migration of the valence basin from one side complex plane to the other side. In $D_{3h}$ symmetry transition state this leads to a drop into the interaction between the cation and the ligand (see the $C_{3v}$ and $D_{3h}$ structure of $[Pb(Cl)_3]^-$, Table 2). The T-shape structure for the TS allows to weaken only two of the lead – ligand interactions. The geometry can also be as square planar complex in which one of the positions is occupied by the valence basin (Figure 3).

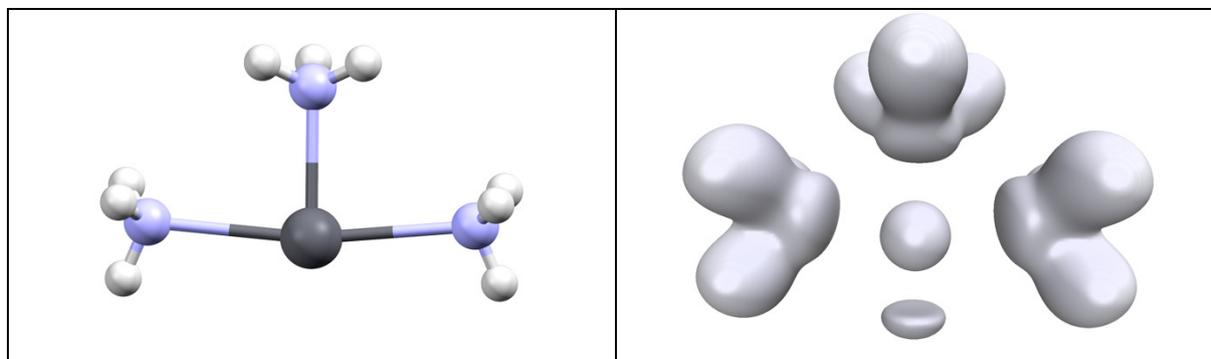

**Figure 7 : Structure (T-shape) of the transition state for the interconversion of the $[Pb(NH_3)_3]^{2+}$ complex (left) and associated ELF analysis (right, isovalue η=0.6).**

| Complex | Structure | ΔG (B3LYP) | ΔG (ωB97XD) |
|---|---|---|---|
| [Pb(H)$_3$]$^-$ | $D_{3h}$ | 53.1 | 50.4 |
| [Pb(Me)$_3$]$^-$ | T-shape | 44.8 | 47.4 |
| [Pb(CN)$_3$]$^-$ | T-shape | 27.0 | 28.7 |
| [Pb(OH)$_3$]$^-$ | T-shape | 23.6 | 25.4 |
| [Pb(NH$_2$)$_3$]$^-$ | T-shape | 20.7 | 22.0 |
| [Pb(OMe)$_3$]$^-$ | T-shape | 19.6 | 21.7 |
| [Pb(F)$_3$]$^-$ | $D_{3h}$ | 16.0 | 17.2 |
| [Pb(SMe)$_3$]$^-$ | T-shape | 15.2 | 16.5 |
| [Pb(SH)$_3$]$^-$ | T-shape | 14.4 | 14.0 |
| [Pb(NH$_3$)$_3$]$^{2+}$ | T-shape | 12.2 | 13.5 |
| [Pb(Cl)$_3$]$^-$ | $D_{3h}$ | 10.7 | 11.5 |
| [Pb(HCN)$_3$]$^{2+}$ | T-shape | 7.9 | 8.3 |
| [Pb(CO)$_3$]$^{2+}$ | T-shape | 6.4 | 6.4 |
| [Pb(OH$_2$)$_3$]$^{2+}$ | T-shape | 5.2 | 5.7 |

**Table 4: General shape and interconverion barriers (in kcal/mol) with B3LYP and ωB97XD functionals.**

To explore further the relationship between the interconversion barriers (Table 4) and the topological properties of the complexes, we plot the barrier values against N(Pb), the electronic population of V(Pb), or d(Pb) (**Figure** 8). To be consistent, the topological analysis we redone on the ωB97XD wavefunction and the barrier values computed at the same level was used for the analysis. The relationship between d(Pb) and the barrier is quite poor though roughly a high barrier value is associated to a larger d(Pb). However, three groups of ligand can be identified. The neutral ligand (OH$_2$, HCN, CO or NH$_3$) are associated to low barriers and short d(Pb) distances. A second group consisting of H$^-$ and H3C$^-$ exhibits both high barrier and large d(Pb). The other anions are intermediate but a significant gap (roughly 20 kcal/mol) distinguish them with the second group. The link between N(Pb) and the barrier is more significant, though the three groups already identified are still valid. This supports our analysis of a direct link between the σ-donor capacity of the ligand, the topology of the valence basin of lead and the interconversion barrier.

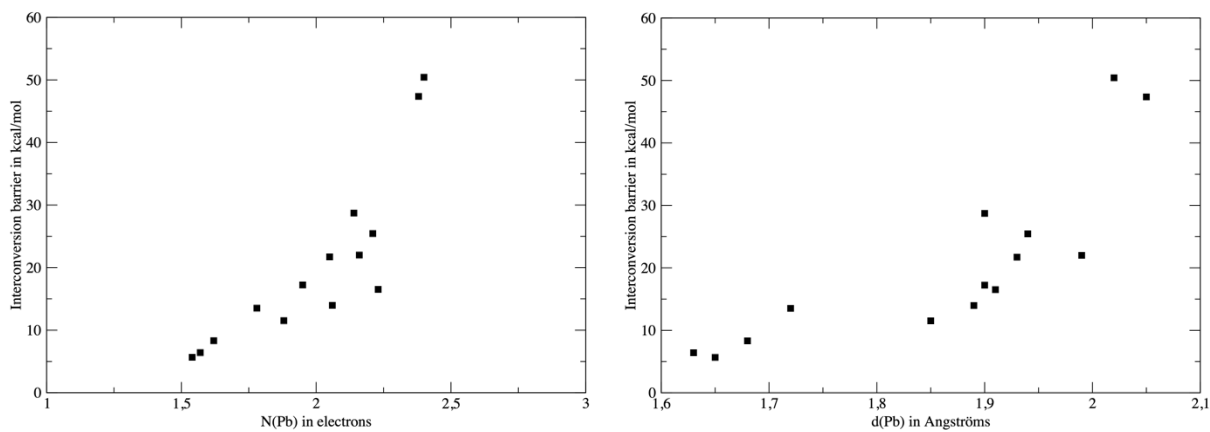

**Figure 8 :** values of the interconversion barrier against N(Pb), left or d(Pb) right, values are computed ate DFT level of theory with ωB97XD functional.

**IV Conclusion**

In this contribution, we have studied the nature of the lead valence basins and especially what we call lone pair in previous studies. Through molecular orbital analysis, NBO and NPA analyses and study of molecular contribution to ELF basin, we have shown that the structures of the lead complexes were only govern by the 6s and 6p subshells. No involvement of the d orbitals has been observed: the depopulation of the 5d orbitals is small and does not seem to be ligand dependant according to NBO analysis and charge transfer to the empty 6d is negligible. The 5d subshell is too low in energy and too contracted to interact with the valence orbital of the ligands. Similarly, the 6d orbitals are too high in energy and does not contribute. The complexation of $Pb^{2+}$ can be explained through interaction with the $6s^2$ electrons and the accepting 6p orbitals.

The number of ELF attractors depends on the nature of the $Pb^{2+}$ ligands, the less donor ones (essentially the neutral ones as $OH_2$) generate three attractors, whereas the stronger σ-donor generate only one attractor. It is thus difficult to reduce the presence of the valence basin to a hybridized $sp^3$ lone pair, due to the weak contribution of the p orbitals to this lone pair as already shown by Shimoni.[11]

The existence of this valence basin has structural and dynamic consequences. Generated by the promotion of 6s electrons to the 6p orbitals and by the σ-donation from the ligand, its volume and population will increase with the strength of the Pb-L interaction. Simultaneously, the distribution of the ligand will be governed by the trans effect. This means that if a strong σ-donor ligand binds to $Pb^{2+}$, the position in trans (considering an octahedron) will be destabilized as we observed on our attempt to optimize the $[Pb(CN)_3(CO)_3]^-$ complex.

Also, this imposes that a second σ-donor ligand will bind preferentially in cis position. This is due to the polarization of the valence basin of lead, which is repelled trans to the ligand and strongly weakens the electrophilicity of the cation in this position in the sense that locally the electronic density increases. In $[Pb(L)_3]^q$ complexes, this can even lead to local nucleophilicity as illustrated by the preference for an interaction with the H atom of HCl of the $[Pb(CN)_3]^-$ complex. This opens the way for the design of new families of specific ligand for $Pb^{2+}$ having a positive pole in their structure.

Other consequences concern the dynamical effect of this lone pair. To our knowledge, no simulation has been done on $[Pb(L)_4]^{2-}$ complexes. If energy minima have been found, the question of their dynamic stability has still not been answered. Indeed, the strong destabilization of the two anions located in trans position opens the way of fast exchange with solvent molecule, to form a $[Pb(L)_3]^-$ complex plus a solvated $L^-$. Another consequence is the configuration stability of $[Pb(L)_3]^-$ complexes. If the cation binds to three different ligands, especially organic ones, the interconversion barrier may be high enough to block the configuration opening the way of the synthesis of enantio pure compounds.